# Coherent phonons in a Bi$_2$Se$_3$ film generated by an intense single-cycle THz pulse

A. A. Melnikov[1,*], K. N. Boldyrev[1], Yu. G. Selivanov[2], V. P. Martovitskii[2], S. V. Chekalin[1], and E. A. Ryabov[1]

[1]*Institute for Spectroscopy RAS, Fizicheskaya 5, Troitsk, Moscow, 108840 Russia*
[2]*P. N. Lebedev Physics Institute RAS, Moscow, 119991 Russia*

*e-mail: melnikov@isan.troitsk.ru

We report an observation of coherent phonons of $E_g^1$, $E_u^1$, $A_{1g}^1$, and $E_g^2$ symmetry generated in a single-crystal film of Bi$_2$Se$_3$ by an intense single-cycle THz pulse. The atomic vibrations reveal themselves through periodic modulation of the refractive index of the film. The largest signal is detected at the frequency of 4.05 THz that corresponds to the $E_g^2$ mode. The generation of $E_g^2$ phonons is interpreted as resonant excitation of the Raman mode by the second harmonic of THz-driven nonlinear $E_u^1$ oscillator, the fundamental frequency of which (2.05 THz) is approximately half that of $E_g^2$. The origin of nonlinearity in this case is cubic lattice anharmonicity, while generation of $E_g^1$ (1.1 THz) and $A_{1g}^1$ (2.25 THz) phonons is a manifestation of quartic anharmonicity enhanced by the occasional combination relations between phonon frequencies in Bi$_2$Se$_3$.

**Introduction.**

The development of laser sources of ultrashort pulses in mid-infrared and terahertz domains has made it possible to study highly nonequilibrium states of matter previously inaccessible in experiments with pulses at visible wavelengths. Mid-IR and THz pulses can selectively excite low-energy degrees of freedom reducing concomitant population of higher lying electronic states and avoiding generation of a hot thermalized electronic distribution. The exposure of a sample to intense low-frequency pulses tuned into resonance with a specific mode results in generation of a high-amplitude coherent wave packet on the timescales shorter than characteristic inverse rates of most relaxation processes. Such excited states are therefore convenient to use as models for studies of nonlinear effects, phase transitions, and interactions between rotational, vibrational, spin and electronic degrees of freedom in their "pure" form. The number of applications of such an approach is growing fast [1, 2]. Among them are the control of electron-phonon interactions by intense mid-IR pulses [3-5] and photoinduced superconductivity [6, 7], THz driving of magnon resonances [8-10], control of magnetic [11, 12], electronic [13, 14] and structural [15-18] properties via coherent pumping of phonon resonances.

In the present work we study the ultrafast response of crystalline $Bi_2Se_3$ to intense THz pulses. This solid belongs to a class of bismuth and antimony chalcogenides, which are objects of intense scientific research in two large areas. First, these crystals demonstrate unique electronic properties predicted for 3D topological insulators [19]. They are narrow bandgap insulators in the bulk, while the surface hosts Dirac-like electronic states with linear dispersion. These electrons are topologically protected from scattering and demonstrate spin-momentum locking. Second, a number of crystals from this family ($Bi_2Te_{3-x}Se_x$) demonstrate a large thermoelectric effect [20]. The figure of merit that is used to evaluate their efficiency is inversely proportional to the thermal conductivity [21]. The latter, in turn, is small when the scattering of longitudinal acoustic phonons is strong. In $Bi_2Te_{3-x}Se_x$ materials and several other thermoelectrics (like $Pb_{1-x}Sn_xTe$ [22] and SnSe [21, 23]) the efficient scattering channel is provided by TO phonon modes that are considerably softened due to lattice anharmonicity [20]. It is well known that reflectivity spectra of $Bi_2Se_3$ crystals measured at room temperature contain a relatively intense broad line that appears in the range 60-70 cm$^{-1}$ and is associated with the TO phonon mode of $E_u^1$ symmetry (reported frequencies are e.g. 65 cm$^{-1}$ in [24] or 61 cm$^{-1}$ in [25]). Therefore, it would be interesting to excite this mode by intense resonant THz radiation and look for nonlinear structural effects.

In the present work we performed such an experiment using nearly single-cycle THz pulses with high peak electric field strength. We found that the optical response of the $Bi_2Se_3$ crystal to the THz pulse contains oscillations not only at the frequency of $E_u^1$ mode that was directly excited, but also at the frequencies of three Raman active modes of $E_g^1$, $A_{1g}^1$, and $E_g^2$ symmetry. Recently, observations of THz-induced oscillating signals were reported for $Bi_2Se_3$, $Bi_2Te_3$, and $Sb_2Te_3$ in [26] and [27]. The appearance of oscillations at the frequency of $E_g^2$ Raman active mode was interpreted as vibrationally induced symmetry breaking in the bulk and sum-frequency Raman scattering respectively. We argue here that the coherent generation of Raman active modes by an intense THz pulse is a result of lattice anharmonicity and multi-phonon processes somewhat similar to the ultrafast ionic Raman scattering proposed in [15].

**Experimental details.**

The sample used in our experiments was a 24 nm thick $Bi_2Se_3$ film grown on a (111)-oriented $BaF_2$ substrate by molecular beam epitaxy. To reduce the exposure of $Bi_2Se_3$ to ambient air the film was protected by a layer of $BaF_2$ with a thickness of 28 nm. The details of sample preparation and notes on structural quality of the $Bi_2Se_3$ film are provided in the Supplemental Material [28].

Terahertz pulses with a duration of about 1 ps were generated in a lithium niobate crystal using optical rectification of femtosecond laser pulses with tilted fronts (a description of this method can be found elsewhere [29]). The THz generation stage was fed by 60 fs laser pulses at 800 nm with an average beam power in the range 0.15 - 1.2 W (i.e. 0.15 - 1.2 mJ per pulse at 1 kHz repetition rate). THz radiation was focused into the sample using parabolic mirrors and the resulting peak electric field reached values of about 1 MV/cm. Dry nitrogen was continuously pumped through the setup to reduce absorption by water molecules (except measurements with spectrally selective excitation that were done in air).

In the experiments we detected transient anisotropic and isotropic changes of transmittance of the sample caused by pump THz pulses. For that purpose we used weak 60 fs probe pulses with a central wavelength of 800 nm. For anisotropic detection the polarization of probe pulses before the sample was set to 45° relative to the vertical polarization of pump THz pulses. Both pump and probe beams were almost normal to the sample surface. Passing through the sample excited by THz radiation the probe pulses experienced a small rotation of polarization, which was detected by measuring the intensities of two orthogonal polarization components of the probe beam $I_x$ and $I_y$ using a Wollaston prism and a pair of photodiodes placed after the sample. The dependence of the value $1-I_x/I_y$ on the probe delay time represented the anisotropic photoinduced response of $Bi_2Se_3$. To detect isotropic changes of transmittance we used the standard pump-probe layout, in which the first photodiode measured the intensity of probe pulses, while the second photodiode registered reference pulses that did not pass through the sample. As a result, the relative change of transmittance $\Delta T/T$ as a function of the probe delay time was obtained.

**Results and discussion.**

As follows from the reflectance spectra of our $Bi_2Se_3/BaF_2$ sample (Fig. S4 of the Supplemental Material [28]), the absorption line that corresponds to the $E_u^1$ phonon mode is located near 67 cm$^{-1}$ (2 THz). The amplitude of this line is independent on the polarization of light due to the $D_{3d}^5$ symmetry of bulk $Bi_2Se_3$. A typical waveform and the corresponding spectrum of the pump THz pulse are shown in Fig. 1. One can see that near 2 THz the spectral amplitude of radiation is large (about half the maximal amplitude near 1 THz) and therefore $E_u^1$ mode can be coherently driven by the electric field irrespective of sample orientation.

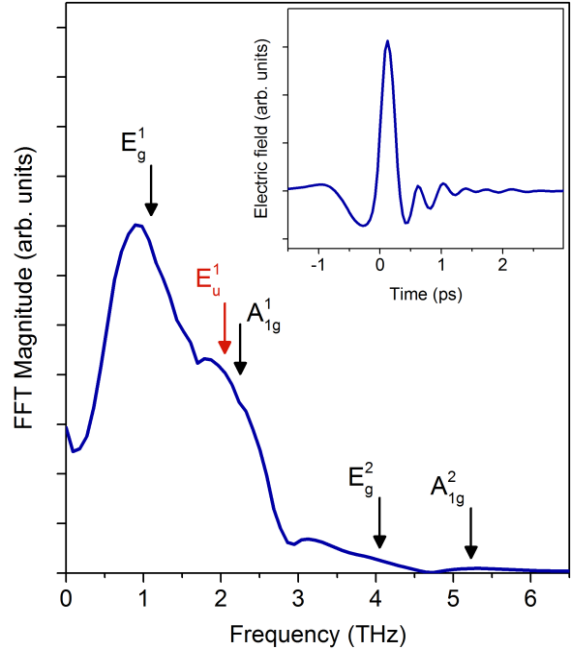

Fig. 1. Time profile of the pump THz pulse (in the inset) and its spectrum obtained using fast Fourier transform (FFT). Arrows show the position of relevant phonon modes of $Bi_2Se_3$ on the frequency scale.

The transient changes of transmittance of the sample (isotropic detection) and transient polarization rotation (anisotropic detection) are illustrated by Fig. 2. In each case the THz-induced response consists of a monotonic and an oscillating component. The former can be ascribed to the relaxation of electrons excited by the pump THz pulse, while the latter is due to coherent motion of atoms of $Bi_2Se_3$ that is commonly referred to as coherent phonons [30-39]. We treat the contribution of $BaF_2$ to the measured traces as negligible, since reference experiments with clean $BaF_2$ substrates revealed no signal above the noise level.

The discussion of the monotonic component of the signals is beyond the scope of the present work. However, we note a considerable difference in electronic relaxation rates that can be extracted from the measured traces. The decay part of the isotropic response can be well fitted by a single exponent with a time constant of $\tau \approx 2.5$ ps. At the same time the monotonic part of the anisotropic response demonstrates approximately two-exponential decay with $\tau_1 \approx 0.5$ ps and $\tau_2 \approx 1.5$ ps, while the amplitude of the longer living component is rather small.

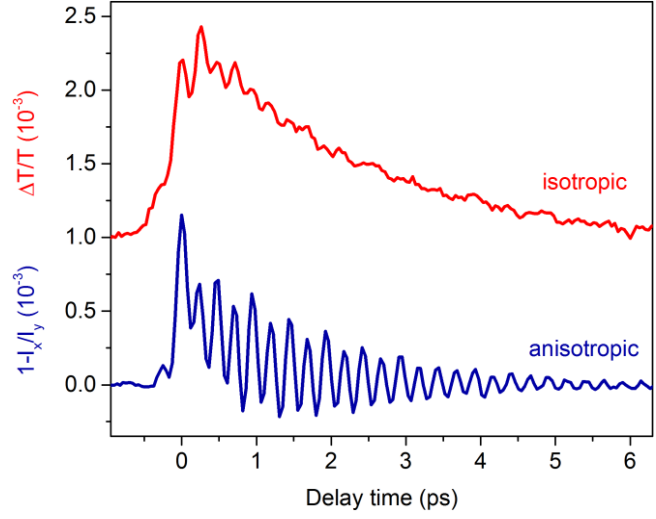

Fig. 2. Response of the Bi$_2$Se$_3$ film to the pump THz pulse measured using isotropic (ΔT/T denotes relative change of transmittance, the curve is shifted upwards by 10$^{-3}$) and anisotropic detection.

In order to find frequencies contained in the oscillating component we applied fast Fourier transform to the time domain data. To reduce noise and artifacts in the frequency domain we measured decay traces several times using different sampling intervals (i.e. steps of the delay line). FFT spectra of individual traces were then averaged to produce spectra shown in Fig. 3. It is possible to distinguish clearly three peaks in the spectrum of the anisotropic signal (1.1, 2.05, and 4.05 THz) and two peaks in the spectrum of the isotropic signal (2.25 and 4.05 THz). We assign these spectral lines to the $E_g^1$ (1.1 THz), $E_u^1$ (2.05 THz), $A_{1g}^1$ (2.25 THz), and $E_g^2$ (4.05 THz) phonon modes of Bi$_2$Se$_3$ [24, 40].

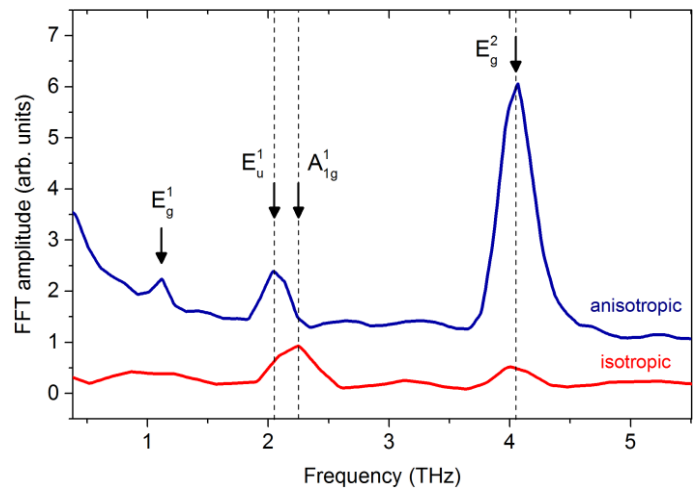

Fig. 3. Spectra of oscillations in the anisotropic and isotropic THz-induced response of Bi$_2$Se$_3$. The anisotropic spectrum was shifted upwards to facilitate comparison. Arrows illustrate attribution of the observed spectral lines to phonon modes of the crystal.

It should be noted that, ideally, the anisotropic detection setup is able to register only coherently excited non-fully symmetric modes ($E_g$ and $E_u^1$ modes in our case), while the isotropic one detects only fully symmetric modes ($A_g$ modes). Fully symmetric coherent phonons induce isotropic variations of reflectivity and/or transmittance and the polarization state of the probe pulse remains unchanged. Therefore, the oscillations are cancelled out upon anisotropic detection. Non-fully symmetric oscillations, in contrast, alter the refractive index of the sample in such a way that the polarization of the probe pulse rotates and its intensity is unaffected. The origin of the residual signal at the $E_g^2$ frequency in the isotropic response in Fig. 3 can be associated with polarization sensitive optical elements (in particular, metallic mirrors) that guide probe beam to the photodiode.

We start the analysis of the observed frequency components from the interpretation of the signal at 2.05 THz in the anisotropic response. Since the corresponding $E_u^1$ mode is IR active, damped oscillations of polarization that follow its coherent excitation by the THz pulse can reveal themselves through the linear electro-optic (Pockels) effect and be detected in our experimental layout. However, Pockels effect in bulk $Bi_2Se_3$ is forbidden by symmetry due to the presence of the center of inversion. It is nevertheless allowed at the surface of the crystal, which has $C_{3v}$ (3m) symmetry. According to this model, if the sample is non-excited, the thin layer of $Bi_2Se_3$ in the vicinity of the $Bi_2Se_3/BaF_2$ interface can be treated as a uniaxial nonlinear crystal. Excitation of coherent atomic $E_u^1$ oscillations is equivalent to the application of the external electric field perpendicular to the optical axis of this nonlinear crystal. As a result, the index ellipsoid of the latter is deformed and a certain difference $\Delta n$ appears between refractive indices experienced by vertically and horizontally polarized components of the probe light. This $\Delta n$ is then indirectly detected in the anisotropic layout. Thus, the amplitude of the registered oscillations at the $E_u^1$ frequency $A(E_u^1)$ must demonstrate linear dependence on the electric field strength of the THz pulse $E_{THz}$, provided the latter is not so strong that effects of saturated absorption are significant. Moreover, the dependence of $A(E_u^1)$ on sample orientation must be in accordance with the $C_{3v}$ symmetry of $Bi_2Se_3$ surface. We performed model calculations (see the details in the Supplemental Material, Section 2 [28]) and found that in case of the electro-optic origin of the signal at 2.05 THz its amplitude $A(E_u^1) \propto |\sin 3\varphi|$, where $\varphi$ is the sample rotation angle.

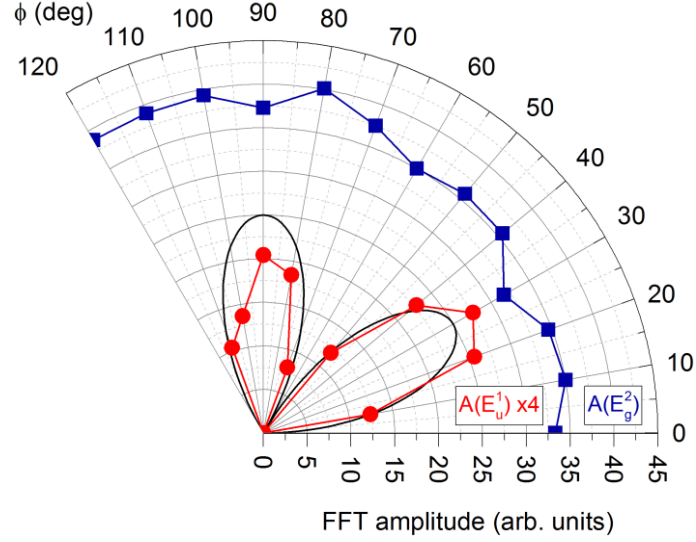

Fig. 4. Dependence of coherent phonon amplitudes $A(E_u^1)$ (multiplied by 4, circles) and $A(E_g^2)$ (squares) on the sample orientation. The angle between THz pump and 800 nm probe polarizations was kept constant (45°). The solid line is the $r = a|\sin 3\varphi|$ function with an arbitrary chosen amplitude $a$.

We measured the dependence of $A(E_u^1)$ on sample orientation (for $\varphi$ in the range 0-120°) and on the power $P_{800nm}$ of the 800 nm laser beam that was fed into the THz generation stage. The results are shown in Fig. 4 and Fig. 5. It was found that $A(E_g^2)$ is almost independent on $\varphi$, while $A(E_u^1)$ is strongly modulated and approximately follows the predicted $A(E_u^1) \propto |\sin 3\varphi|$ dependence. The invariance of $E_g^2$ amplitude is the consequence of $D_{3d}^5$ symmetry of the bulk $Bi_2Se_3$ as soon as for such crystals the intensities of Raman lines do not depend on the orientation of crystallographic axes in the geometry used in our measurements [41]. As follows from Fig. 5(a), the dependence of $A(E_u^1)$ on laser power can be relatively well fitted by a square root function: $A(E_u^1) \propto \sqrt{P_{800nm}}$. In the optical rectification technique the intensity of THz pulses $I_{THz}$ is proportional to $P_{800nm}$ in a rather wide range. Since $E_{THz} \propto \sqrt{I_{THz}}$, the observed root dependence of $A(E_u^1)$ on $P_{800nm}$ implies its linear variation with $E_{THz}$. Thus, relying on the obtained data, we can conclude that $E_u^1$ mode is excited across the whole volume of the $Bi_2Se_3$ film, and the corresponding signal at 2.05 THz originates from the $Bi_2Se_3/BaF_2$ interfaces through electro-optic modulation of the refractive index associated with surface electronic states of $Bi_2Se_3$.

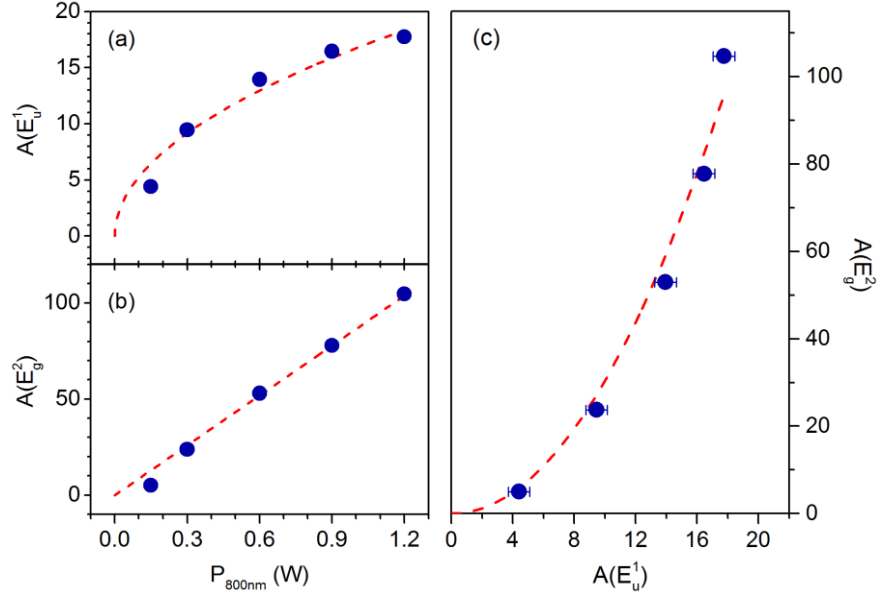

Fig. 5. (a), (b) - The dependence of coherent phonon amplitudes $A(E_u^1)$ and $A(E_g^2)$ on the power of the laser beam fed into the THz generation stage ($P_{800nm}$). (c) - The dependence of $A(E_g^2)$ on $A(E_u^1)$. Dashed lines represent fits of the experimental data: square root in panel (a), linear in (b), and parabolic in (c).

In contrast to the $E_u^1$ mode, $E_g^1$, $A_{1g}^1$, and $E_g^2$ phonon modes of Bi$_2$Se$_3$ are Raman but not IR active and coherent oscillations of atoms at the corresponding frequencies cannot be excited directly through resonant linear absorption of the THz radiation (as, for example, in tellurium [42]). Coherent excitation of Raman phonons in solids is commonly described using ISRS [43], TSRS [44] or DECP [45] theories but those mechanisms are inefficient in Bi$_2$Se$_3$ because the duration of THz pulses in our case is about 1 ps that is almost equal to (for $E_g^1$ mode) or considerably larger (for other modes) than the periods of the phonon modes.

In order to interpret the observed phenomenon, the features of coherent oscillations at 2.05 and 4.05 THz ($E_u^1$ and $E_g^2$ modes) should be considered, in particular, the relative variation of their amplitudes $A(E_u^1)$ and $A(E_g^2)$ with pump THz electric field. Indeed, we found that while $E_u^1$ coherent amplitude increases linearly with the THz electric field strength, the $E_g^2$ one demonstrates quadratic growth (i.e. linear in $P_{800nm}$, Fig. 5(b)). The point measured at $P_{800nm}$ = 0.15 W deviates from the trend, but the same deviation can be seen in Fig. 5(a) for the first point of the $A(E_u^1)$ power dependence. Nevertheless, if we plot $A(E_g^2)$ as a function of $A(E_u^1)$, the obtained data set can be fitted by a parabola within the experimental error (Fig. 5(c)).

This nonlinear relation between $A(E_u^1)$ and $A(E_g^2)$ implies that some second-order process should be responsible for the generation of coherent $E_g^2$ phonons. A straightforward way to interpret the observations is to suppose that this process is the three-phonon interaction due to lattice anharmonicity, when two $E_u^1$ phonons annihilate producing one $E_g^2$ phonon. This process is allowed by symmetry, because the interaction can be described by the term $\propto q_I^2 q_R$ in the oscillator energy (where $q_I$ and $q_R$ are coordinates of the infrared active and Raman active oscillators respectively), and for the $D_{3d}$ point group $E_u \otimes E_u = A_{1g} + A_{2g} + E_g$. In addition, energy is conserved in this process, as soon as the frequency of $E_g^2$ mode is approximately twice the frequency of $E_u^1$ mode. We note here that recently lattice anharmonicity was invoked to explain THz excitation of Raman active modes in CdWO$_4$ [46].

In case of Bi$_2$Se$_3$ the excitation of coherent Raman phonons can be modelled using the approach similar to the one proposed in [15]. Let $q_I(t)$, $q_R(t)$ and $m_I$, $m_R$ be coordinates and masses of the infrared active (*I*) and Raman active (*R*) oscillators. Then the corresponding equations of motion will have the form

$$m_I \ddot{q}_I = -\frac{\partial}{\partial q_I}\left(U - B q_I E_{THz}(t)\right), \quad (1)$$

$$m_R \ddot{q}_R = -\frac{\partial}{\partial q_R} U, \quad (2)$$

where

$$U = \frac{1}{2} m_I \omega_I^2 q_I^2 + \frac{1}{2} m_R \omega_R^2 q_R^2 - A q_I^2 q_R \quad (3)$$

is the potential energy of oscillators including the anharmonic interaction energy $-A q_I^2 q_R$. For the sake of simplicity we neglect the damping. The term $-B q_I E_{THz}(t)$ accounts for the interaction of the IR active oscillator with the electric field of the pump THz pulse, and *B* is a constant that characterizes the strength of this interaction. After differentiation we obtain

$$m_I \ddot{q}_I + m_I \omega_I^2 q_I = 2A q_I q_R + B E_{THz}(t), \quad (4)$$

$$m_R \ddot{q}_R + m_R \omega_R^2 q_R = A q_I^2. \quad (5)$$

We consider the amplitude of THz-driven IR active oscillator to be much larger than that of the Raman active oscillator ($q_I \gg q_R$) and neglect the term $2A q_I q_R$. Then, making the substitution $q_I = \sqrt{m_I m_R} Q_I$, $q_R = m_I Q_R$, and $B/m_I^{3/2} m_R^{1/2} \to B$ we get

$$\ddot{Q}_I + \omega_I^2 Q_I = B E_{THz}(t), \quad (6)$$

$$\ddot{Q}_R + \omega_R^2 Q_R = A Q_I^2. \tag{7}$$

Since damping is neglected, we put to a certain approximation that after the resonant action of the THz pulse $Q_I(t) \propto \cos(\omega_I t)$. Then the driving term $AQ_I^2$ contains two contributions: the first one is constant, while the second one is $\propto \cos(2\omega_I t)$. The quasiconstant driving force (in fact, it represents a convolution of a step function and the envelope of the THz pulse) is considered in [15] as a source of transient lattice displacement by analogy with nonlinear optical rectification. This force, however, can excite $Q_R$ only if $\omega_I \gg \omega_R$, which is not the case for $Bi_2Se_3$, as its optical phonon modes lie in the THz range and have frequencies of the same order. Nevertheless, coherent Raman phonons can be resonantly excited by the second term $\propto \cos(2\omega_I t)$ if $\omega_R = 2\omega_I$, and in $Bi_2Se_3$ this condition is fulfilled, as soon as $\omega(E_g^2) \approx 2\omega(E_u^1)$.

In order to better illustrate this model we solved numerically the equations of motion for the IR active and Raman oscillators using the real temporal profile of pump THz pulses as the driving force. For that purpose we introduced damping into Eqs. 6 and 7 so that

$$\ddot{Q}_I + \gamma_I \dot{Q}_I + \omega_I^2 Q_I = B E_{THz}(t), \tag{8}$$

$$\ddot{Q}_R + \gamma_R \dot{Q}_R + \omega_R^2 Q_R = A Q_I^2, \tag{9}$$

where $\omega_I = 2$ THz, $\omega_R = 4$ THz, constants $A$ and $B$ were chosen arbitrarily, while the ratios $\omega_I/\gamma_I \approx 1.5$, $\omega_R/\gamma_R \approx 0.9$ - so that the damping of oscillations in the calculated traces was close to the damping observed in the experiment. The results are presented in Fig. 6. Panel (a) contains the THz waveform measured by electro-optic detection (The same curve is shown in the inset to Fig. 1). These values were substituted into the right side of Eq. 8 in order to solve it numerically. The obtained solution $Q_I(t)$ is plotted in Fig. 6(b). One can see that after the excitation $Q_I(t)$ represents exponentially decaying oscillations, characteristic of a damped oscillator, and the contribution of transient processes is negligible. We used this $Q_I(t)$ to calculate the trajectory of the Raman oscillator $Q_R(t)$ from Eq. 9. As follows from Fig. 6(c), the Raman oscillator demonstrates a delayed ~ 1 ps response that matches the lifetime of the driving force $Q_I^2(t)$ and is approximately five times longer than the duration of a hypothetic nonlinear force proportional to the squared electric field of the THz pulse. In order to compare this model to the experimental results we tried to extract oscillating waveforms that correspond to the coherent atomic motions at 2 THz and 4 THz from the anisotropic response. For that purpose we convolved the latter with wavelet functions at carrier frequencies of 2 THz and 4 THz, while the duration of the wavelets was 600 fs and 400 fs respectively. Such values were chosen in an attempt to keep both temporal and frequency resolution. These wavelets are shown in Fig. 7 along with the results of the convolution.

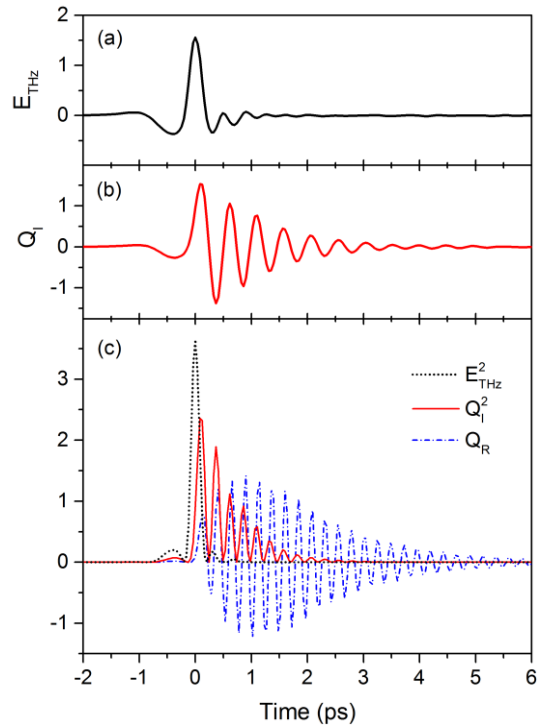

Fig. 6. (a) - Electric field of the pump THz pulse substituted as a driving force into Eq. 8. (b) - Trajectory of the IR active oscillator obtained solving Eq. 8. (c) - Trajectory of the Raman active oscillator driven by $Q_I^2$, the solution of Eq. 9 (dash-dotted curve). The dotted curve represents the squared electric field of the THz pulse as a reference.

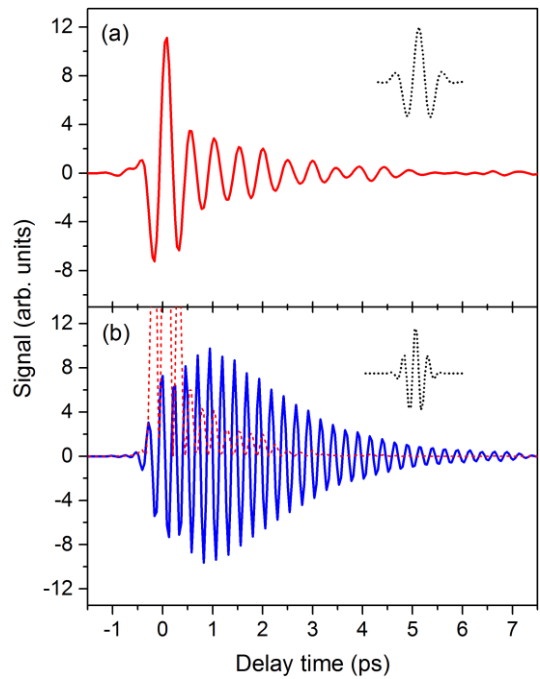

Fig. 7. (a) - Convolution of the experimental anisotropic response of $Bi_2Se_3$ with the wavelet at 2 THz having a duration of ~ 600 fs. (b) - Convolution of the same anisotropic signal with the wavelet at 4 THz having a duration of ~ 400 fs. The dashed line is the squared trace from panel (a). The wavelets are shown as dotted curves.

Figure 7(b) shows that coherent oscillations at 4 THz are launched with a delay and reach the maximal amplitude approximately 1 ps after excitation, similar to the trajectory of the Raman oscillator presented in Fig. 6(c). However, the trace at 2 THz in Fig. 7(a) demonstrates a considerable difference from its counterpart $Q_I(t)$ in Fig. 6(b) in the vicinity of zero delay time. The difference is due to the rather intense wavelet-like signal that results from the convolution of the 2 THz wavelet with the zero time spike in the anisotropic response (see Fig. 2). This spike not necessarily corresponds to the motion of atoms of Bi$_2$Se$_3$ and can be caused by some short lived purely electronic response of the sample to the THz pulse (e.g. via Kerr effect). When we plot the squared 2 THz trace along with the 4 THz trace as shown in Fig. 7(b), the resulting picture is rather similar to the one obtained above by solving equations of motion for corresponding oscillators (Fig. 6(c)), if one neglects this intense signal near zero time delay. Thus, the results of numerical modeling and time-frequency analysis of the experimental data speak in favor of our assumption that highly excited anharmonic $E_u^1$ lattice vibrations provide the force that generates coherent Raman active phonons through anharmonic coupling.

We note that the observed gradual picosecond growth of the amplitude of $E_g^2$ coherent phonons contradicts the model of sum frequency Raman scattering proposed in [27], according to which the nature of generation process is displacive and the oscillations start almost immediately upon the arrival of the pump THz pulse without any delay.

We emphasize here also the difference between our observations and the results reported in [26]. Similar to our work the authors of [26] detected oscillations at frequencies near 2 THz and 4 THz in the photoinduced response of Bi$_2$Se$_3$ using, however, a different detection technique - time-resolved second harmonic generation (SHG). Important is the fact that according to our data the damping of the lower frequency mode is comparable or stronger than the damping of the higher frequency mode. Moreover, as shown above, the corresponding wavepackets are shifted in time domain relative to one another. The opposite is true for the oscillations observed in [26]. The 2 THz mode has a longer lifetime than the 4 THz one, while oscillations at both frequencies start almost simultaneously upon THz excitation. Thus, it is possible to assume that different phenomena are observed in [26] and in our work. In our case the oscillations at 4 THz are caused by coherent $E_g^2$ phonons through anisotropic modulation of the refractive index. In [26] the signal at 4 THz is "derived" from the oscillations of polarization at 2 THz via the second-order nonlinear optical effect, while coherent $E_g^2$ phonons cannot be observed in the SHG response due to symmetry restrictions [38].

We finish the analysis of $E_u^1$ and $E_g^2$ coherent phonons by checking the resonant character of the oscillatory THz response. For that purpose we performed an additional experiment, in which we filtered the pump THz radiation to selectively attenuate the spectral band near 2 THz. We used a crystal of $PrFe_3(BO_3)_4$ that was at hand, however, it is possible to use any other suitable filter. The reflectance spectrum of $PrFe_3(BO_3)_4$ is shown in the inset to Fig. 8(b) [47]. At ~ 1.7 THz the reflectance experiences a sharp drop and remains low near 2 THz.

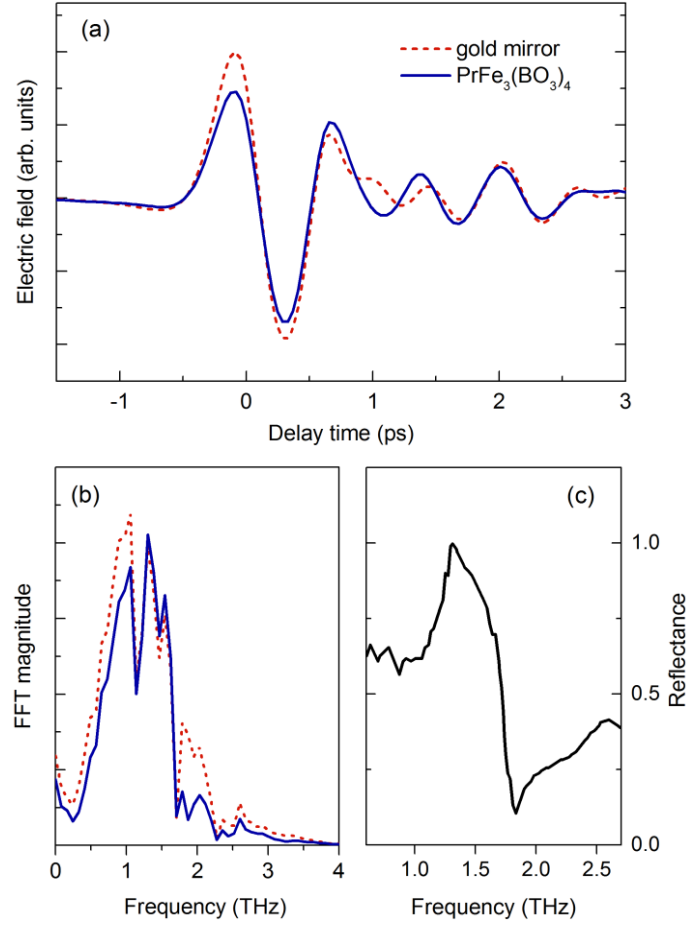

Fig. 8. (a) - Electric field of the pump THz pulse reflected from a gold mirror (dashed line) and from the $PrFe_3(BO_3)_4$ crystal (solid). (b) - Spectra of the THz waveforms shown in (a). (c) - Reflectance spectrum of $PrFe_3(BO_3)_4$ [47]. Measurements were made under ambient conditions.

If we substitute one of the gold mirrors in the pump channel by this crystal, the amplitude of the THz pulse will be reduced by a factor of 1.24, while the spectral amplitude near 2 THz - by ~ 2. At the same time, as follows from Fig. 9, the amplitude of $E_g^2$ coherent phonons will decrease by a factor of ~ 4 and not 1.5 as it would be if the response near 4 THz was proportional to the squared electric field strength. This result shows that the second-order lattice response of $Bi_2Se_3$ is determined by the spectral amplitude of pump THz radiation near 2 THz, or, in other words, requires a resonance with the $E_u^1$ phonon mode.

Finally, we discuss the origin of $E_g^1$ and $A_{1g}^1$ coherent phonons. Excitation of these modes cannot be interpreted taking into account only three-phonon processes by analogy with the generation of $E_g^2$ mode, because energy is not conserved in this case. However, it is reasonable to consider also four-phonon processes (recently it was proposed that quartic anharmonicity causes anomalous hardening of $A_g^1$ mode in $Bi_2Te_3$ [20]). Taking into account the broadening of phonon lines that relaxes to a certain extent the law of energy conservation, it is possible to construct two frequency combinations: $2\omega(E_u^1) \approx \omega(A_{1g}^2) - \omega(E_g^1)$ and $2\omega(E_u^1) \approx 2\omega(A_{1g}^1)$.

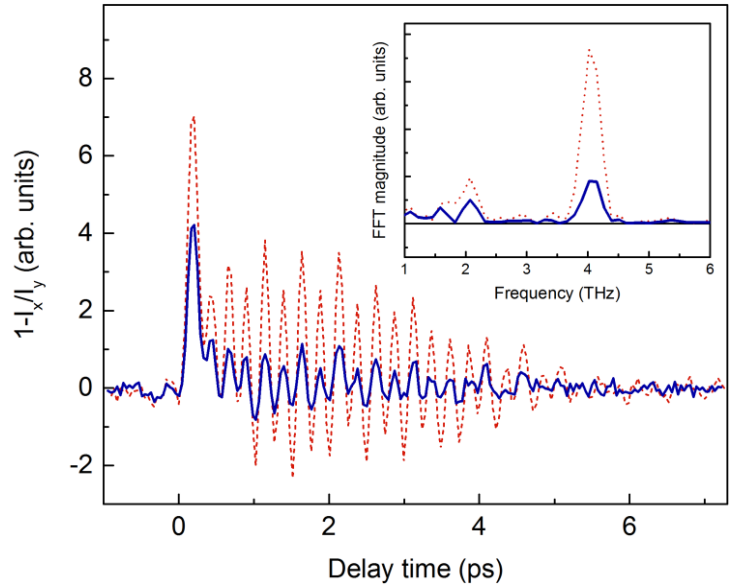

Fig. 9. THz response of $Bi_2Se_3$ measured using THz pulses reflected from a gold mirror (dashed line) and from the $PrFe_3(BO_3)_4$ crystal (solid line). The inset shows spectra of these signals. Measurements were made under ambient conditions.

The first process implies interaction of two $E_u^1$, one $E_g^1$ and one $A_{1g}^2$ phonons, and, as a result of intense excitation of the $E_u^1$ mode, generation of coherent oscillations both at 1.1 THz ($E_g^1$) and 5.23 THz ($A_{1g}^2$) frequencies. So far we were unable to reliably detect the $A_{1g}^2$ oscillations in the THz-induced response of $Bi_2Se_3$ probably because of the relatively long probe pulse. Indeed, its duration was 60 fs - about one third of the $A_{1g}^2$ phonon period, which would lead to additional reduction of the oscillating signal amplitude by a factor of ~ 3 relative to the detection with the ultimate resolution. Another possible reason is that the modulation of the refractive index by $A_{1g}^2$ phonons is weak, as one can conjecture reviewing the results of pump-probe measurements in the visible and infrared domains, where this mode had the smallest amplitude of all detected modes (see e.g. [33]).

Nevertheless, the $E_g^1$ peak is clearly visible in the anisotropic spectra, and we adhere to the model with four-phonon interaction. In this case the anharmonic coupling term in the potential energy will have the form $Aq_I^2 q_{R_1} q_{R_2}$, where indices *1* and *2* refer to the $E_g^1$ and $A_{1g}^2$ modes. And for the equations of motion of the two Raman modes we obtain two coupled equations

$$\ddot{Q}_{R_1} + \omega_{R_1}^2 Q_{R_1} = AQ_I^2 Q_{R_2}, \tag{10}$$

$$\ddot{Q}_{R_2} + \omega_{R_2}^2 Q_{R_2} = AQ_I^2 Q_{R_1}. \tag{11}$$

Analogous to the case of $E_g^2$ mode discussed above, the driving terms on the right side of these equations will contain harmonics at combination frequencies of $\omega(A_{1g}^2) - 2\omega(E_u^1)$ and $\omega(E_g^2) + 2\omega(E_u^1)$, and therefore resonantly excite corresponding Raman oscillators.

As for the $A_{1g}^1$ mode, its anharmonic coupling to the $E_u^1$ mode will be represented by the term $Aq_I^2 q_R^2$ (interaction of two $E_u^1$ and two $A_{1g}^1$ phonons). In this case the equation of motion has the form

$$\ddot{Q}_R + \omega_R^2 Q_R = AQ_I^2 Q_R \text{ or} \tag{12}$$

$$\ddot{Q}_R + \left(\omega_R^2 - AQ_I^2\right) Q_R = 0. \tag{13}$$

Again, $AQ_I^2 \propto 1 + \cos 2\omega_I \approx 1 + \cos 2\omega_R$. The constant term causes renormalization of $\omega_R$, while modulation of $\omega_R$ at the frequency of $2\omega_R$ excites the oscillator parametrically.

**Conclusion.**

In summary, we have measured the response of a single crystal Bi$_2$Se$_3$ film to a nearly single-cycle THz pulse. We have found that the intense resonant excitation of the IR active $E_u^1$ mode leads to the generation of $E_g^1$, $A_{1g}^1$, and $E_g^2$ coherent Raman active phonons. We conclude that this effect is caused by three- and four-phonon interactions in the anharmonic crystal lattice of Bi$_2$Se$_3$. The origin of the force that drives Raman modes is the nonlinear motion of $E_u^1$ oscillator. The observed effect differs from conventional Raman scattering, for which the driving term acts only through electronic degrees of freedom. The force is not impulsive like in ISRS, and not displacive like in DECP, but periodic and has a certain duration defined by the THz pulse duration and the damping rate of the driven $E_u^1$ mode.

Coherent $E_g^2$ phonons should also be generated via this process in related crystals of $Bi_2Te_3$ and $Sb_2Te_3$, for which $\omega(E_g^2) \approx 2\omega(E_u^1)$ as in $Bi_2Se_3$. It would be interesting also to detect other THz-induced coherent Raman phonons in the whole family of compounds. Using such control parameters as doping and temperature, it could be possible to investigate the interplay of cubic and quartic lattice anharmonicities in these solids. The approach used here can also be applied for studies of lattice anharmonicities in a broader class of thermoelectrics, many of which have soft TO modes.

## Acknowledgements.

Authors thank A. A. Sokolik for useful discussions. This work was supported by the Ministry of Education and Science of the Russian Federation (project #RFMEFI61316X0054). The experiments were performed using the Unique Scientific Facility "Multipurpose femtosecond spectroscopic complex" of the Institute for Spectroscopy of the Russian Academy of Sciences.

## References.